\pdfoutput=1
\documentclass{JINST}

\newcommand{\bdec }{\mbox{$\beta$~decay}}

\newcommand{\ttwo }{\mbox{$\rm T_2$}}
\newcommand{\PO }{\mbox{\textbf{PartOpt}}}
\title{A scheme for the determination of the magnetic field in the KATRIN main spectrometer }

\author{{A. Osipowicz$^a$\thanks{Corresponding
author.}, U. Rausch$^a$, A. Unru$^a$,  B. Zipfel$^b$ } \\
\llap{$^a$} Department of Electrical Engineering and Information Technology, University of Applied Sciences,
Marquardstr. 35, Fulda, Germany\\
\llap{$^b$}HF-Dept., GSI Helmholtzzentrum für Schwerionenforschung GmbH, Planckstrasse 1, D-Darmstdt, Germany\\ 
 
E-mail: \email{Alexander.Osipowicz@et.hs-fulda.de}}

\abstract{To determine
the magnetic field distribution in the  KATRIN main-spectrometer with magnetic field sensors that are placed outside the main-spectrometer vessel one can utilize the absence of magnetic rotation in main-spectrometer volume. There a scalar magnetic potential $V(\vec{x})$ can be defined that fulfills the Laplace equation. Large numbers of magnetic field values on an outer surface of the main-spectrometer can be sampled by moving and fixed magnetic field sensors. These surface samples are used as boundary values in the relaxation of the Laplace equation for $V(\vec{x})$ and the magnetic field components in the volume.  In a simulation involving  the KATRIN reference solenoid chain, a global magnetic field and an external perturbing solenoid it is shown that with this method the original field can be reconstructed within 2 \%.}


\begin{document}

\section{The KATRIN setup}

\begin{figure}[h]
\centering
\includegraphics[width=.9\textwidth]{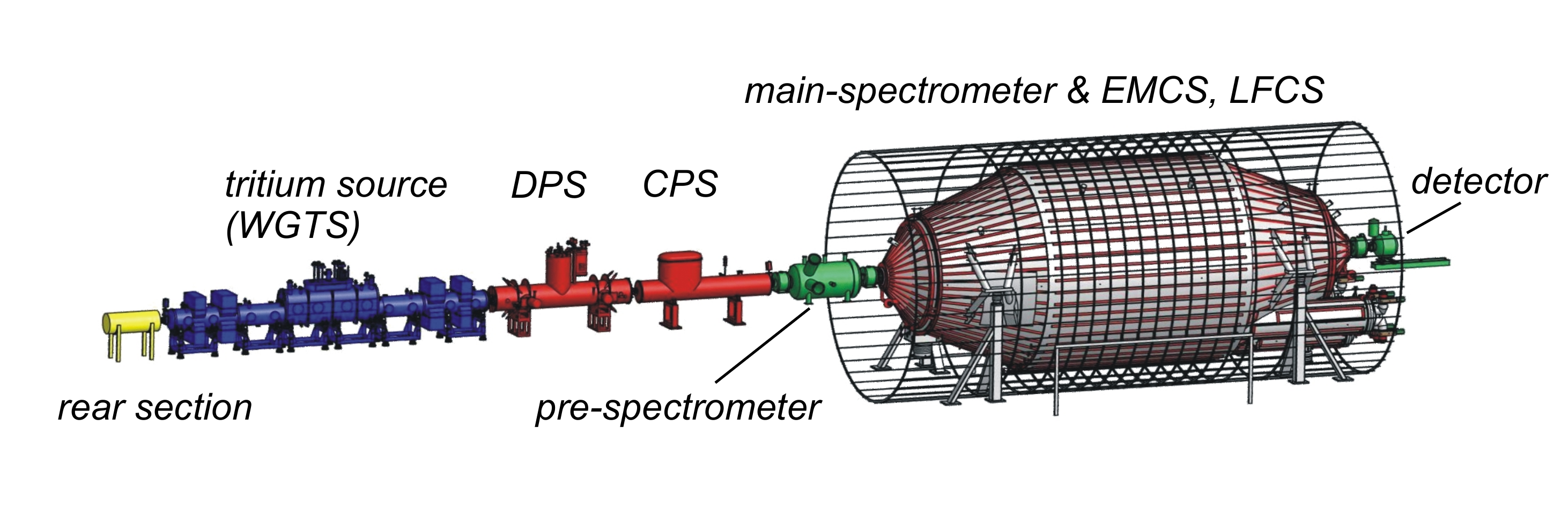}
\caption{\label{fig:KATRIN-all-eng}Schematic view of the KATRIN experiment (total length 70 m) consisting of calibration and monitor rear system, 
with the windowless gaseous \ttwo -source (WGTS), differential pumping (DPS) and cryo-trapping section (SPS), the small pre-spectrometer and the large main spectrometer with the large magnetic coil systems to compensate the earth magnetic field (EMCS) and to shape the magnetic transport flux (LFCS) and lastly the segmented PIN-diode detector.}
\end{figure}

The KArlsruhe TRItium Neutrino  experiment \cite{katrin-loi} (see Fig.\ref{fig:KATRIN-all-eng})  is set up at the Karlsruher Institute of Technology (KIT), Germany. It is designed to measure the mass of the electron anti neutrino in a direct and model-independent way with a
sensitivity of $m_{\nu}=0.2$ eV/c${^2}$ (90\% confidence level) from tritium \bdec \cite{katrin-loi}. KATRIN uses a magnetic transport field  that connects the source and detector in combination with  integrating electrostatic energy filters (MAC-E-spectrometers). 
Conceptual essentials of the MAC-E spectrometer\cite{Pic92,Lob85} are the  magnetic field gradients in pre - and main-spectrometer that adiabatically convert cyclotron energy $E_{cyc}$ into energy $E_p$ parallel to the magnetic field lines and vice versa. 

At the center of the main-spectrometer (MS) in the minimal magnetic field $B_A\approx 3-6$ $\mu$T, a retarding electric field  allows  an integral energy analysis of $E_p$.  The magnetic field  in the analyzing volume defines the magnetic resolution, i.e. the amount of residual cyclotron energy $E_{cyc}$ that can not be analyzed  and thus strongly influences  the resolution function.  Error analysis \cite{VAL} of the influence of uncertainty of the magnetic field in the analyzing plane on the uncertainty of the neutrino mass square $\Delta  m_v^2 $ leads to a relative accuracy of the magnetic field of $\frac{\Delta B}{B}<2.4\%$.  In addition, the  alignment of magnetic field lines plays a crucial role in the production of secondary electrons and electronic background either through penning traps or inner wall contact.

Large coil systems \cite{OSI-1} are arranged around the MS for a) global magnetic field compensation, e.g. earth magnetic field (EMCS) and b) fine tuning of the magnetic transport flux with a set of large circular low field coils (LFCS) mounted coaxially with the MS (see Fig.\ref{fig:KATRIN-all-eng}).
However, possible influences of residual external dipoles, magnetization  in the MS environment by  the  high field solenoids and/or EMCS, LFCS and the correct orientation of the  spectrometer solenoids have to be controlled. Due to the extreme MS vacuum conditions  the installation of magnetic sensors inside the MS  is not possible.

\begin{figure}[h]
\centering
\includegraphics[width=0.6\textwidth]{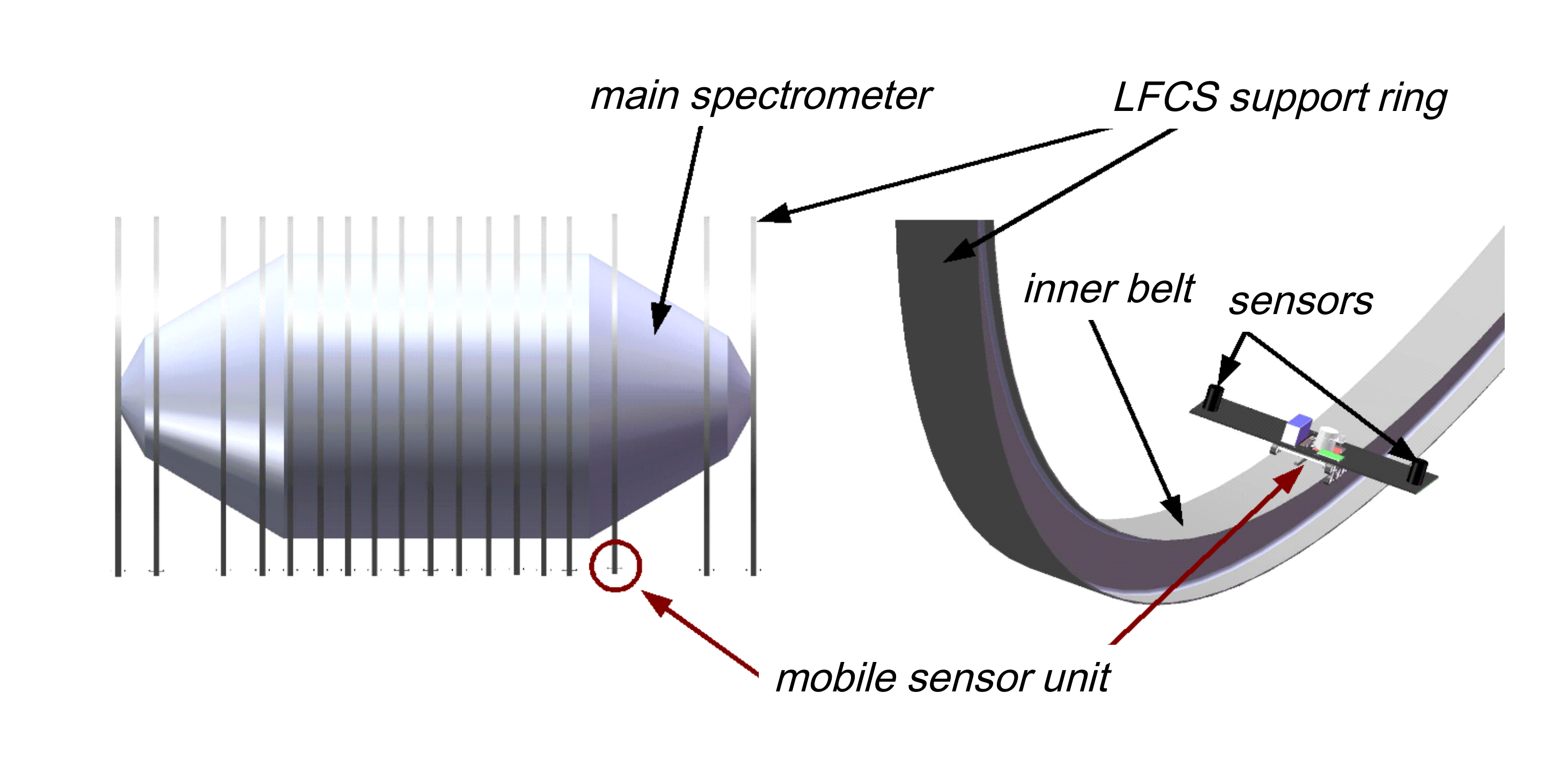}
\caption{\label{fig:MOBS-support-ring}View of the main spectrometer tank with the LFCS ring system. Right: The mobile sensor unit with 2 sensors on the inner belt of a LFCS support ring.}
\end{figure}
 
We therefore propose to determine the magnetic field \textsl{inside} the main spectrometer by taking  magnetic field samples at an  outer surface of the main spectrometer. The sensor network will involve fixed position magnetic sensors and mobile magnetic field sensors \cite{OSI-2,UNRU,LET} which move along the inner belts of the LFCS support structure (see Fig.\ref{fig:MOBS-support-ring}), close to the outer MS surface  but well inside the current lines of the EMCS and LFCS . The magnetic field samples serve as boundary values for the relaxation of the Laplace equation of the scalar magnetic potential $V(\vec{x})$ at the interior of the KATRIN main spectrometer. 

\section{Volume and surface considerations}
\begin{figure}[htbp]
	\centering
		\includegraphics[width=0.8\textwidth]{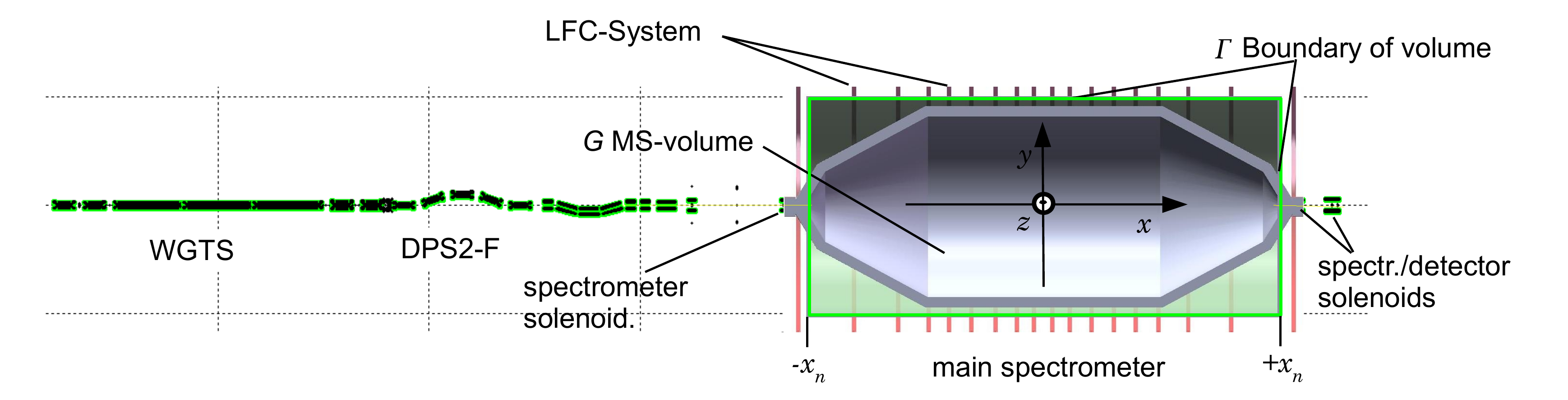}
	\caption{\label{fig:Simulation-1}View of the KATRIN main reference solenoid chain and the main spectrometer area. The cylindrical volume $G$ enclosed by the boundary $\Gamma$. The origin of the coordinate system is located at the geometrical center of the MS and the symmetry point of the LFCS. To ensure that all current leading  elements are outside the boundary $\Gamma$ in $y-$ and $z-$ direction the allowed radius $R_G$ to  $G$ is 5600 mm $ < R_G < $ 6155 mm. The extreme $x-$ values are -11600 mm $=-x_n<x<+x_n=11600$ mm.} 
	\label{fig:PO-simulation-os}
\end{figure}

For a volume  $G$ with surface area $\Gamma$  Amperes equation
\begin{equation}
\vec{\nabla}\times\vec{B}=\mu_0\cdot\left(\vec{J}+\frac{\epsilon_0\cdot\partial\vec{E}}{\partial t}\right)
\label{eq:ampere1}
\end{equation}
can be  simplified to the rotationally free case if the current density $ \vec{J}$ is vanishing ($ J= 0 $) and the electric field $\vec{E}$ is constant ($\partial \vec{E} / \partial t = 0 $). 
\begin{equation}
\vec{\nabla}\times\vec{B}= 0
\label{eq:ampere2}
\end{equation}
For the KATRIN MS the relevant surface $\Gamma$ (see Fig.\ref{fig:Simulation-1}) has to be outside the outer MS surface and inside the current leading elements (LFCS, EMCS, spectrometer solenoids). As the analyzing potential distribution $U(x,y,z)$ inside the MS volume is  constant during KATRIN runtime intervals (and magnetic field sampling time intervals) the electrical fields produced are time independent. Therefore eq. (\ref{eq:ampere2}) can assumed to be valid for the KATRIN MS interior.

Vector analysis  \cite{APO} states for a scalar function $V(\vec{x})$ that: $\vec{\nabla}\times\vec{\nabla}\cdot V(\vec{x})=0$ and one can   identify $V(\vec{x})$ with the magnetic scalar potential. 
\[
\vec{B}=\vec{\nabla}\cdot V(\vec{x})
\]
Utilizing Gauss's law for magnetism $ \vec{\nabla}\cdot\vec{B}=0$
we can write down the \textit{Laplace-equation} (LPE) for $V(\vec{x})$
\begin{equation}
\nabla^{2}V(x,y,z)=0	
\end{equation}
The finite difference method (FDM)  \cite{SCH} is chosen to solve the above equation on a 3 dimensional rectangular grid, because of its well known numerical stability and the manageable coding effort. In the simulation the magnetic field components at a the  boundary representing the normal derivatives $
	\partial V /\partial x= B_x; \: \partial V /\partial y= B_y;\:\partial V / \partial z= B_z $ at  $\Gamma$ can be exported and used in the FD-relaxation as a \textit{von Neumann boundary} values.

\section{Simulation}\label{sec:Simulation}
The usability of the numerical approach is demonstrated in a simulation based on magnetic field values provided by the simulation  package  \PO ~\cite{PO}. The definition of a magnetic scenario (Fig. \ref{fig:PO-robust-sim}) at the KATRIN main spectrometer includes: a) the energized KATRIN reference solenoid chain, b) the energized LFCS  as listed in \cite{GLU1}, c) a  magnetic field  over $G$ with $B_x= 210$ mG, $B_y=35$ mG, $B_z=0$,
d) a small disturbing  magnetic dipole with central induction $ B_{c}= 600$ G adjacent to the main spectrometer.

\begin{figure}[htbp]
\centering
\includegraphics[width=0.9\textwidth]{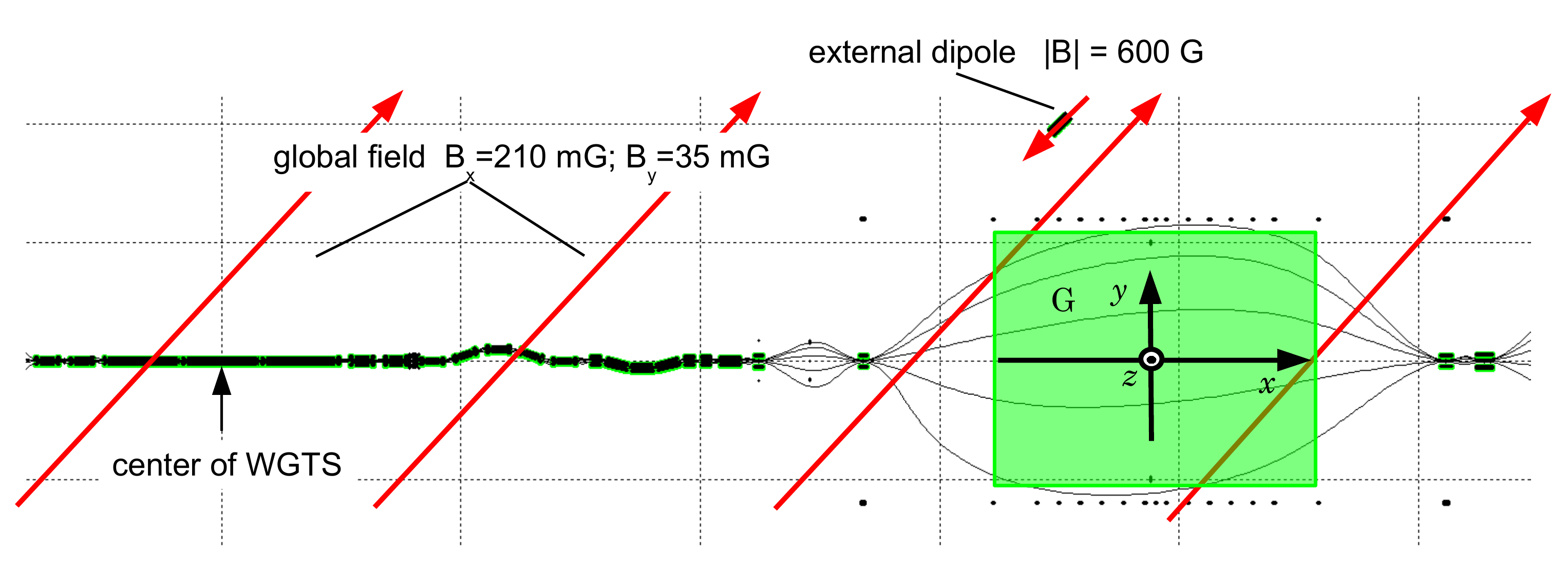}
	\caption{\label{fig:PO-robust-sim} \PO ~view of the Simulation scenario. In volume  $G$ the effective magnetic field is composed of the KATRIN solenoid field, LFCS, an external global field and disturbing external dipole. The perturbed magnetic field lines magnetic field lines have been tracked starting from the center of the WGTS. The extreme field lines indicate the boundary of the $191$ T cm$^2$ nominal magnetic transport flux connecting source and detector.}
	\label{fig:PO-robust-sim}
\end{figure}

The field values $ B_x,B_y, B_z$ along the cylindrical surface of volume $G$ with radius $R_G=6$ m   between $x_{min}=-7.03 $ m $ < x < x_{max}= 6.83 $ m to cover the cylindrical part of the MS are exported in ASCII format. The spacing of the samples in $x$-direction is 0.45 m in agreement with the real $x$-spacing of the sensor positions. In azimuthal direction a $3^{\circ}$ spacing was chosen to get  $120\times 2$ samples (because 2 sensors  are on board) in 15 minutes, the time for one revolution. To simulate sensor error the exported values are randomized according to a Gaussean distribution with a 2\% relative uncertainty. This value was chosen as an upper limit  according to the sensor types used in \cite{OSI-2} .
Due to the cylindrical geometry the surface samples points usually do not coincide with surface mesh points (cut surfaces problem). Therefore the magnetic samples are interpolated to produce values at the regular surface mesh points. The relaxation is performed via a basic $7$ point stencil. The resulting values for the scalar potential and the values for the magnetic field components are generated by deriving $V(\vec{x})$ numerically.

\begin{figure}[h!]
	\centering
		\includegraphics[width=.7\textwidth]{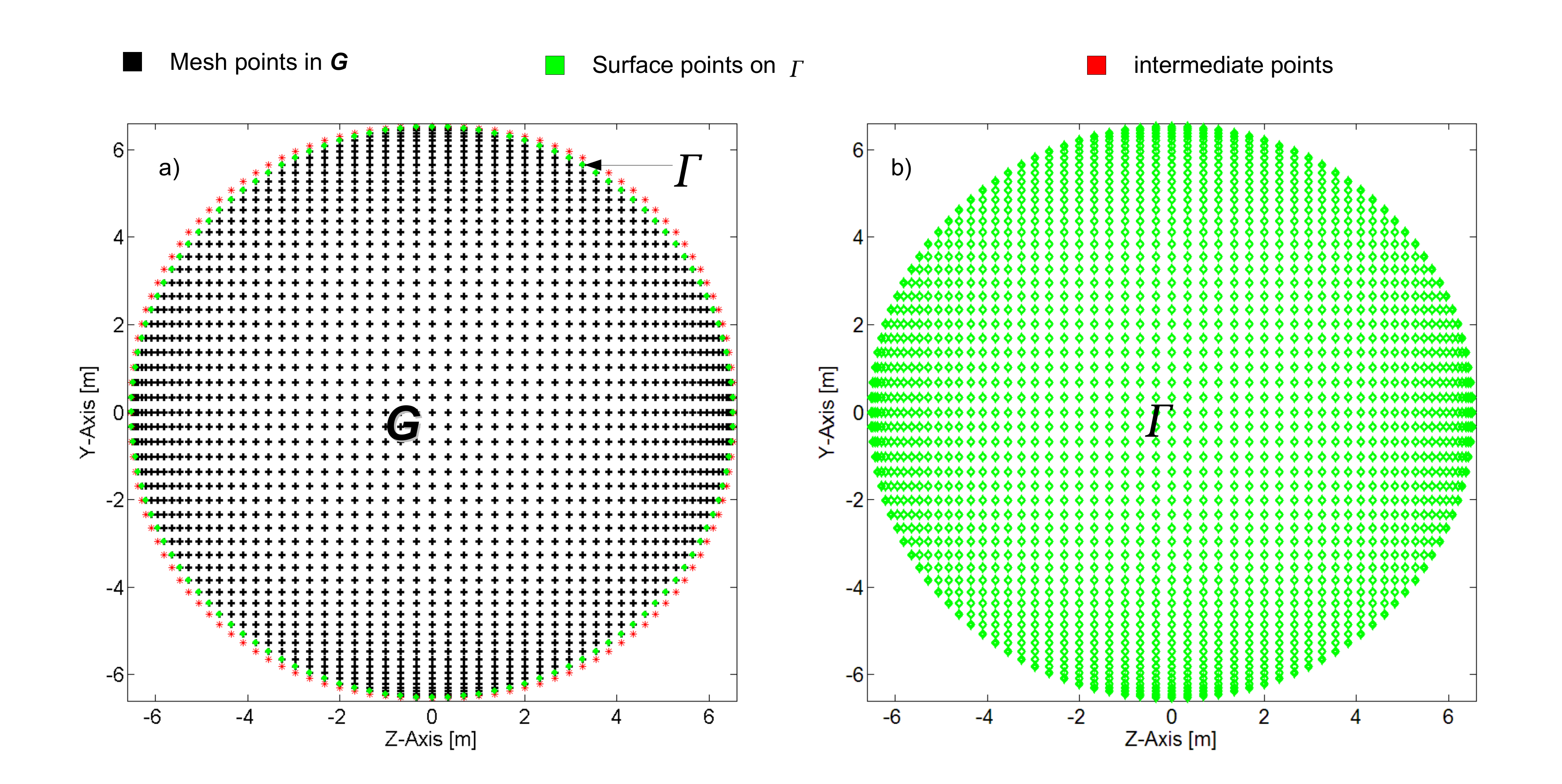}
	\caption{View of the mesh point structure. Left: $G$ in an interval in , $-x_n<x<x_n$, Right: on the surface $\Gamma$, at $x=-x_n, x_n$  }
	\label{fig:masche-os}
\end{figure}
The relaxation code is  written in C. Typically 1400 iterations in 5 minutes on a standard PC are performed to meet the terminating condition that the difference for  $V(0,0,0)$, the magnetic potential at the origin, between successive iterations is $<  0.0002$.

\section{Simulation results}\label{sec:res}
\begin{figure}[h!]
	\centering
		\includegraphics[width=1.1\textwidth]{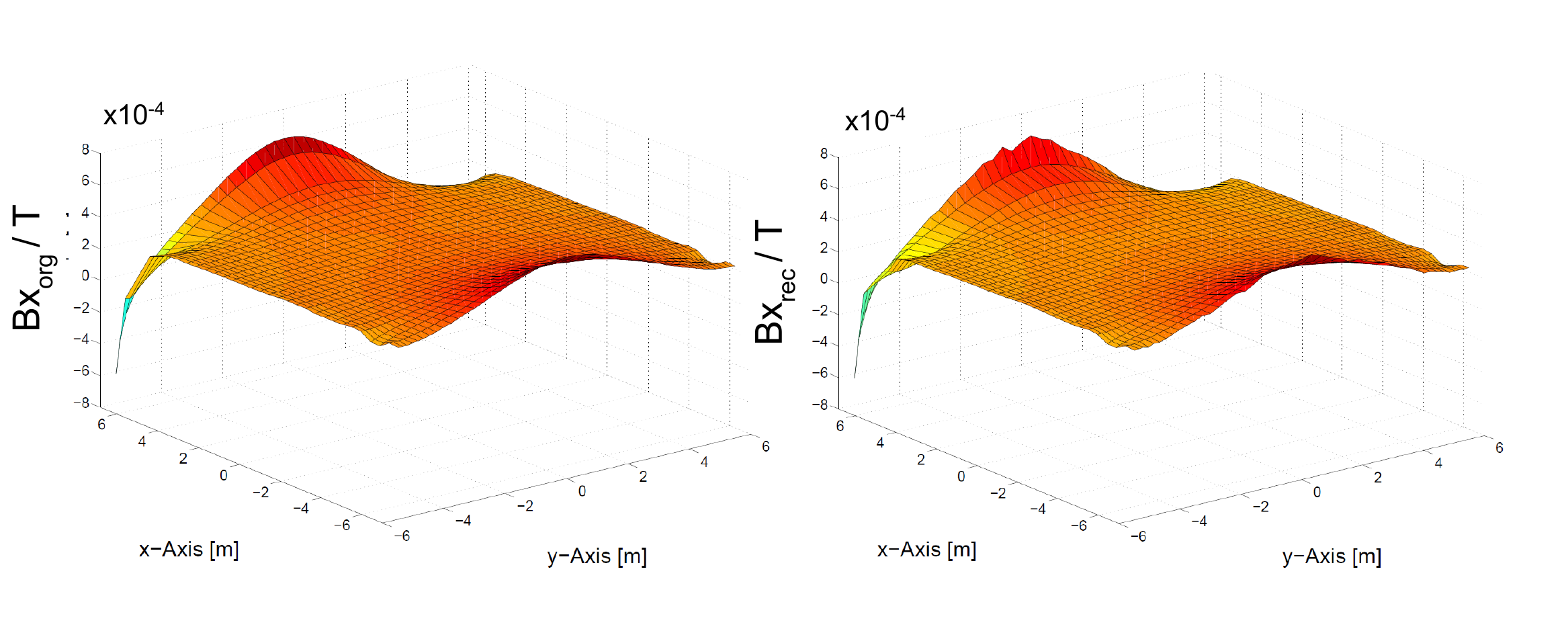}
	\caption{Left: The  original magnetic field component $B_{x_{org}}$ in a  in a $x,y$ plane at $z=2.4994$ m. Right: The reconstructed magnetic field values  $B_{x_{rec}}$ in the same plane.}
	\label{fig:Bx-1}
\end{figure}

\begin{figure}[h!]
	\centering
\includegraphics[width=0.7\textwidth]{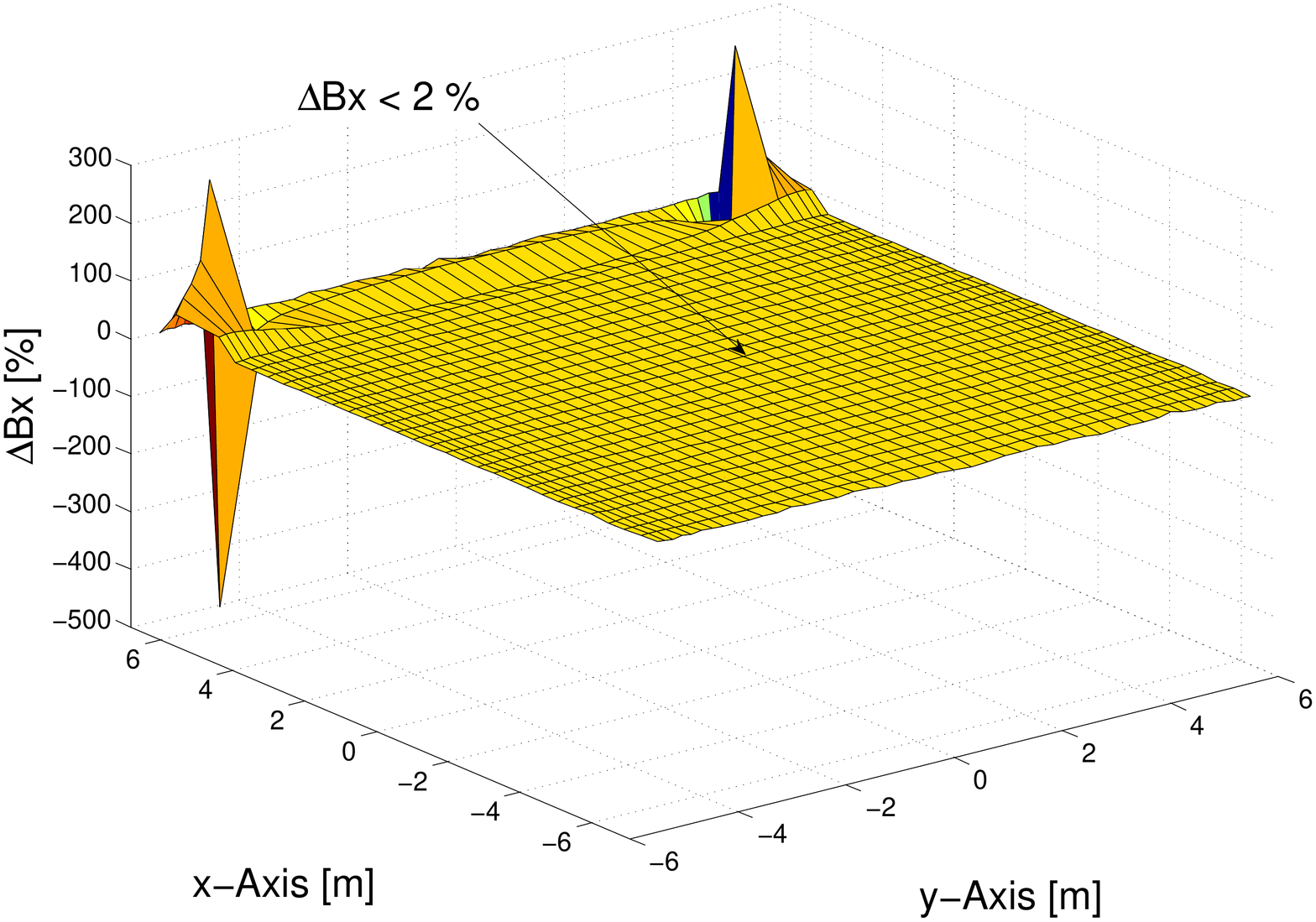}
	\caption{Relative difference  between the original $B_{x_{org}}$ and reconstructed $B_{x_{rec}}$ magnetic field component $\Delta B_x=(B_{x_{org}}-B_{x_{rec}})/B_{x_{org}}$ 
	in a $x,y$ plane with $z=2.4994$ m. The sharp peaks at $x\approx 6$ arise numerically from a division by zero as  $B_{x_{org}}\approx 0 $  in the vicinity of the negatively  charged LFCS coil towards the detector side  as given in \cite{GLU1}. Elsewhere the difference is less than 2\%. }
	\label{fig:deltaBx2}
\end{figure}

\begin{figure}[h!]
	\centering
		\includegraphics[width=1.1\textwidth]{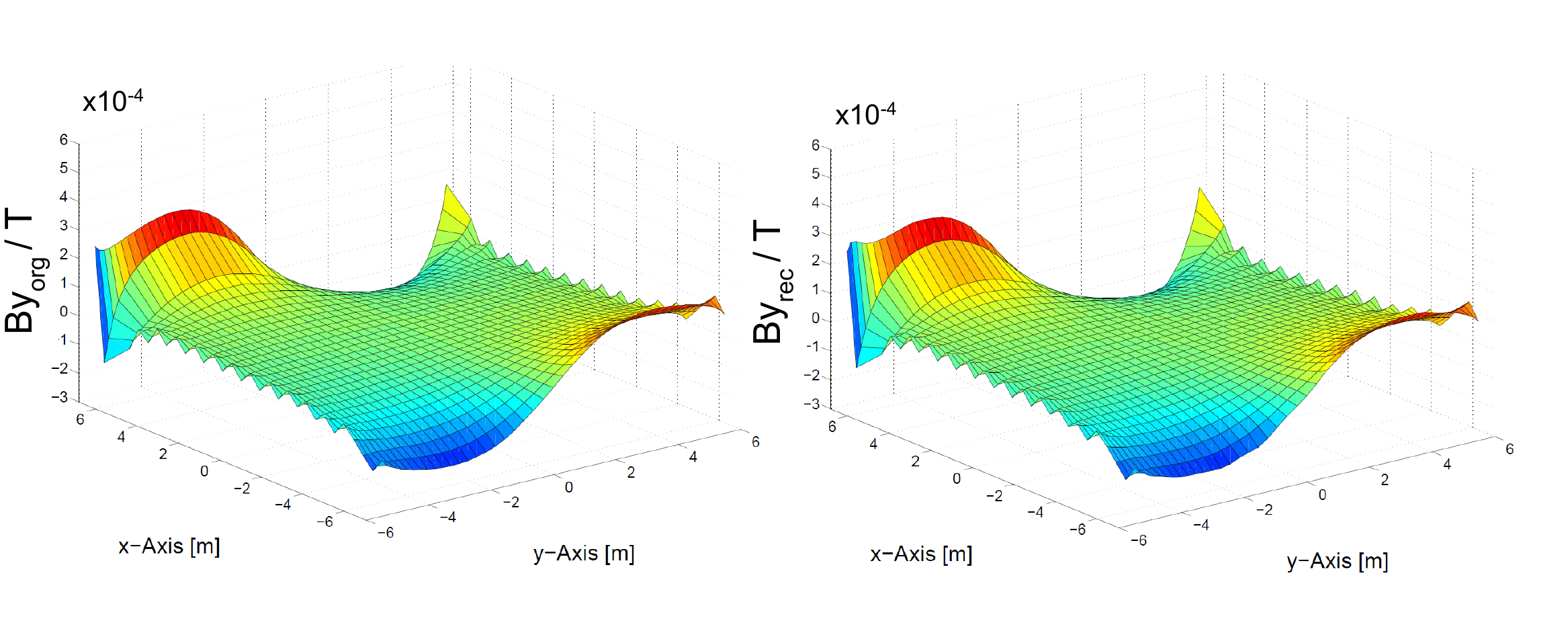}
	\caption{Left: The  original magnetic field component $B_{y_{org}}$ in a  in a $x,y$ plane at $z=2.4994$ m. Right: The reconstructed magnetic field values  $B_{y_{rec}}$ in the same plane. The sawtooth structure at the extreme y-values are due to the close proximity of the energized LFCS Coils.}
	\label{fig:By-1}
\end{figure}

\begin{figure}[h!]
	\centering
		\includegraphics[width=1.0\textwidth]{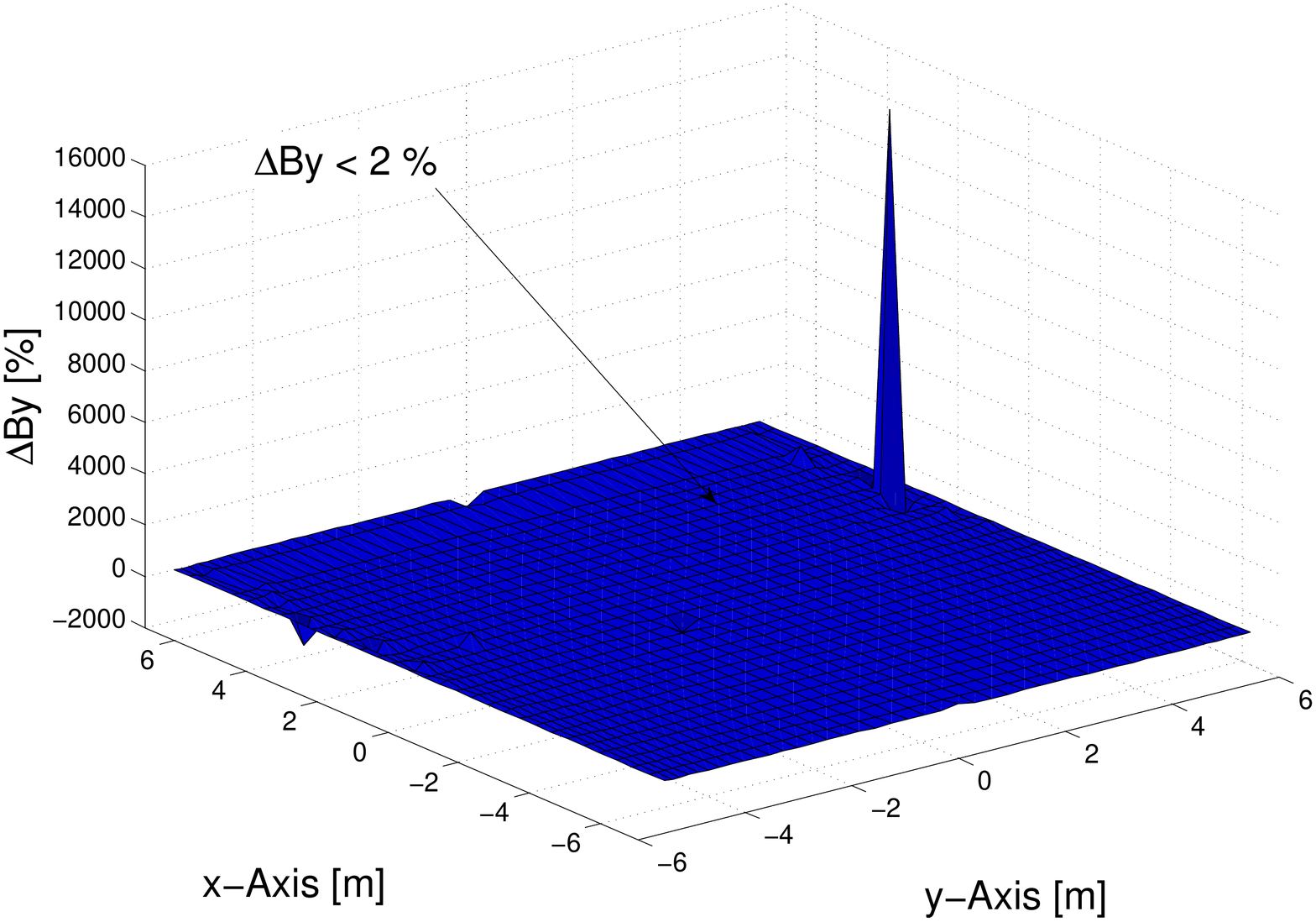}
	\caption{Relative difference  between the original $B_{y_{org}}$ and reconstructed $B_{y_{rec}}$ magnetic field component $\Delta B_y=(B_{y_{org}}-B_{y_{rec}})/B_{y_{org}}$ in a $x,y$ plane with $z=2,4994$ m. The sharp peaks arise numerically from a division by zero as  $B_{y_{org}}\approx 0 $.}
	\label{fig:deltaBy2}
\end{figure}

\begin{figure}[h!]
	\centering
		\includegraphics[width=1.1\textwidth]{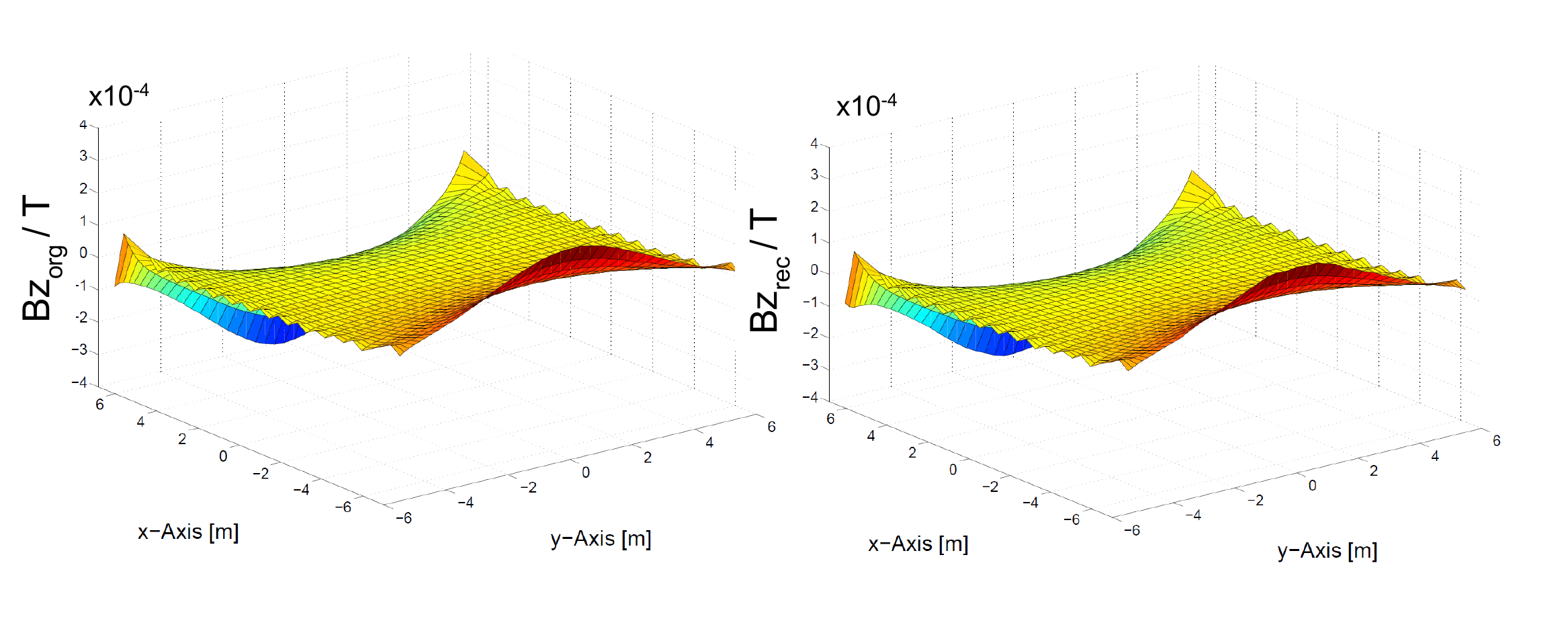}
	\caption{Left: The  original magnetic field component $B_{z_{org}}$ in a  in a $x,y$ plane at $z=2.4994$ m. Right: The reconstructed magnetic field values  $B_{z_{rec}}$ in the same plane. The sawtooth structure at the extreme y-values are due to the close proximity of the energized LFCS Coils.}
	\label{fig:Bz-1}
\end{figure}

\begin{figure}[h!]
	\centering
\includegraphics[width=1.\textwidth]{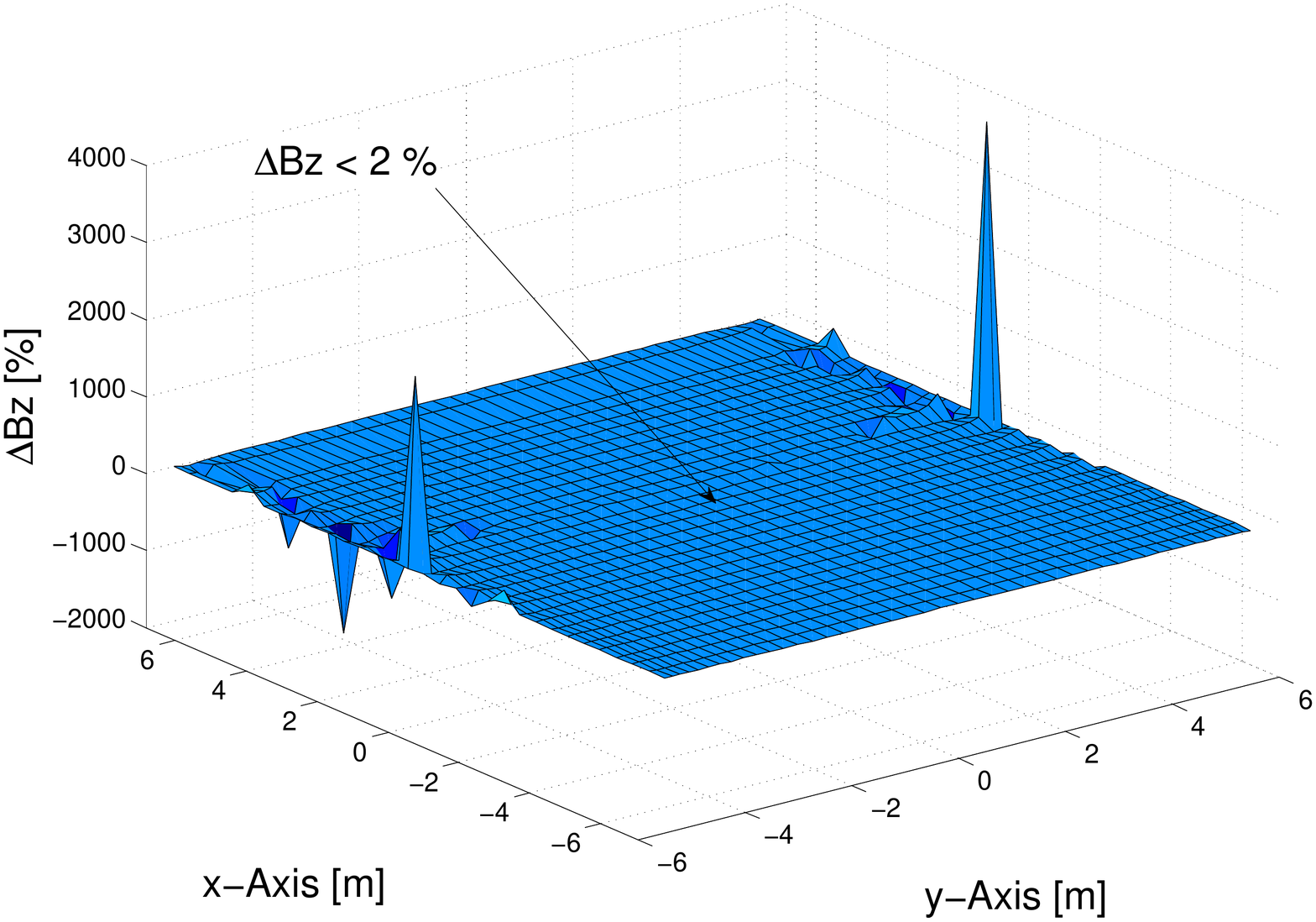}
	\caption{Relative difference  between the original $B_{z_{org}}$ and reconstructed $B_{z_{rec}}$ magnetic field component $\Delta B_z=(B_{z_{org}}-B_{z_{rec}})/B_{z_{org}}$ in a $x,y$ plane with $z=2,4994$ m. The sharp peaks at extreme $y$ -values  arise numerically from a division by zero as  $B_{z_{org}}\approx 0 $.}
	\label{fig:deltaBz2}
\end{figure}

The results of the simulation is displayed as magnetic field components  in geometric planes with given coordinates within the main spectrometer.  Figs.: \ref{fig:Bx-1},\ref{fig:By-1},\ref{fig:Bz-1} show the  original \PO ~ magnetic $B_{org}$   and the reconstructed magnetic field  $B_{rec}$ components for a randomly chosen $x,y$ plane at $z=2.4994$ m. The relative differences $\Delta B$ are displayed in  Fig.: \ref{fig:deltaBx2},\ref{fig:deltaBy2},\ref{fig:deltaBz2},. 



\begin{figure}[h!]
	\centering
\includegraphics[width=1.0\textwidth]{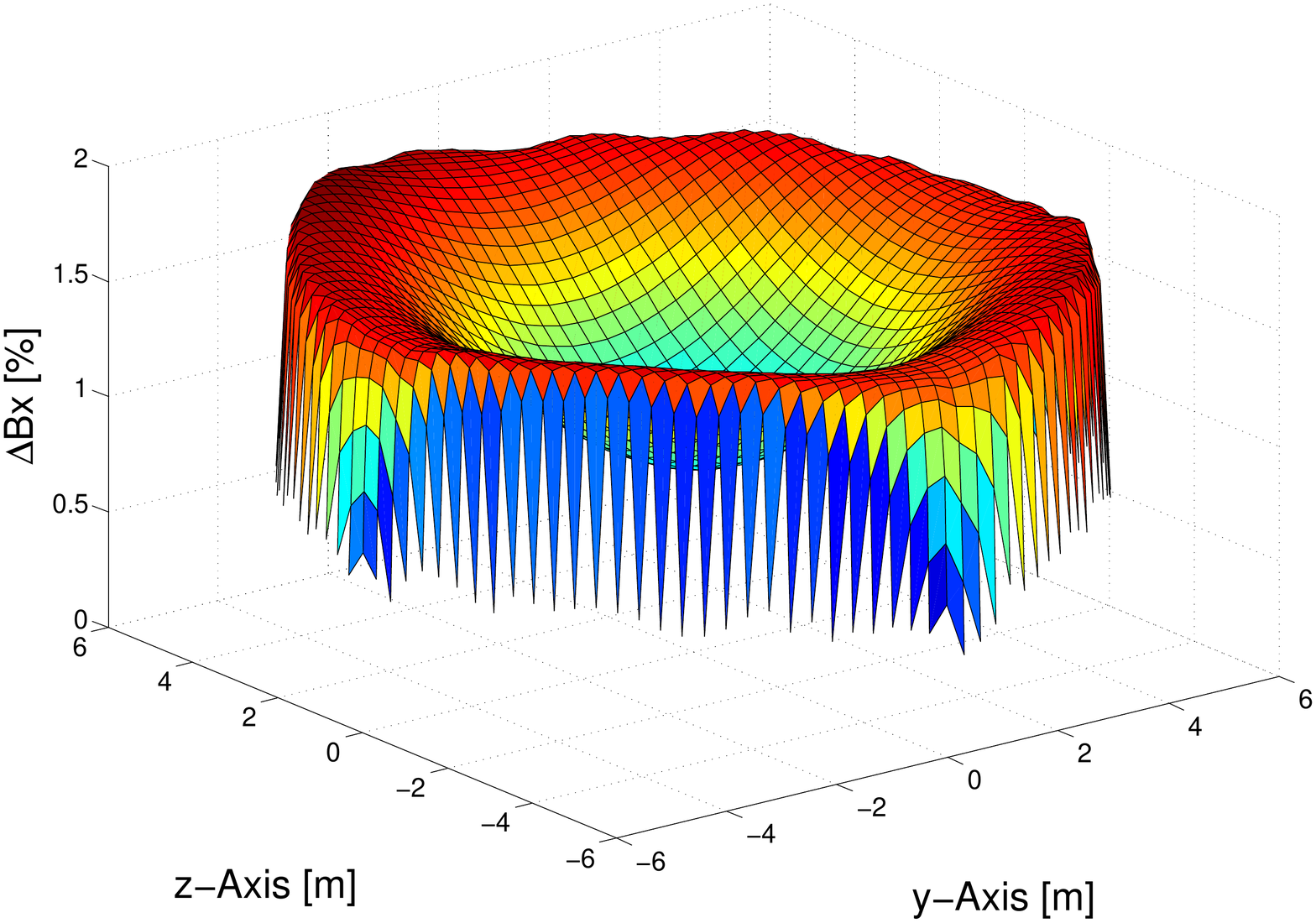}
	\caption{Relative difference  between the original $B_{x_{org}}$ and reconstructed $B_{x_{rec}}$ magnetic field component $\Delta B_x=(B_{x_{org}}-B_{x_{rec}})/B_{x_{org}}$ 
	in a $z,y$ plane with $x=2.475$ m. The sawtooth structure  at the fringes is due to the vanishing $B-x$-component at large radii. }
	\label{fig:deltaBx-1}
\end{figure}

Results with similar precision can be found in all areas of the inner 
volume.Fig. \ref{fig:deltaBx-1} shows the the relative difference $\Delta B_x$ for for a $y,z$-plane at $x=2.475$ m.

\section{Summary and Outlook}\label{sec:summary}
In a simulation it is shown that with a large number of magnetic field samples  taken close to the KATRIN main-spectrometer surface and inside the current leading elements of the LFSC -, EMCS system and spectrometer solenoids it is possible to determine the magnetic field profile inside the spectrometer at least within a 2\% precision. With  better numerical techniques (e.g. stencils involving more meshpoints, interpolation routines with more supporting points) and longer computer relaxation times  an increase in  precission is possible.
Also the number and  distribution of the sampling positions on the surface can in the case of the mobile sensor units be varied to achieve better results.\\
As the front face (at $-x_{n}$)  and the end face (at $+x_{n}$) of the cylindric volume still intersect the KATRIN MS volume no samples can be taken there. However, the magnetic field of these surfaces is predominantly given by the spectrometer solenoids which can be modeled numerically to produce calculated  field values. These models can be controlled by fixed position magnetic field sensors close to the relevant surfaces.\\
Unlike in a simulation, where the magnetic field components are per se given according to the chosen coordinate system, the magnetic field sensors in KATRIN environment have to be aligned according to the KATRIN global coordinate system. In the case of moving sensor units moving on the inner rails of the LFCS structure as proposed in \cite{OSI-2} this requires information about position and inclination along the track.

\vspace{0.5cm}

\noindent\textbf{\sffamily{AKNOWLEDGMENTS}}

\vspace{0.5cm}
The authors wish to express gratitude to the group for
Experimental Techniques of the Institute for Nuclear Physics
(IK) at KIT for highly efficient and competent support. 
Furthermore, we wish to thank Prof. Dr. E. W. Otten, Mainz University and Prof. Dr. Ch. Weinheimer, Münster University for helpful discussions and support.
In addition, we like to thank the University of Applied Sciences, Fulda and the Fachbereich Elektrotechnik und Informationstechnik, for the enduring support for this work.

This work has been funded by the German Ministry
for Education and Research under the Project codes
05A11REA, 05A08RE1.

\end{document}